\documentclass[twocolumn,prl,aps,amsfonts,amssymb,amsmath,showpacs]{revtex4-1}
\usepackage{graphicx}
\usepackage{dcolumn}
\usepackage{mathtools}
\usepackage{color}
\usepackage{braket}
\usepackage{hyperref}
\graphicspath{{figures/}}
\usepackage{soul}
\setkeys{Gin}{width=\linewidth}

\begin{document}

\title{Negative spin exchange in a multielectron quantum dot}

\author{Frederico Martins$^{1,*}$, Filip K. Malinowski$^{1,*}$, Peter D. Nissen$^{1}$, Saeed Fallahi$^{2}$, Geoffrey~C.~Gardner$^{2,3}$, Michael J. Manfra$^{4,5}$, Charles M. Marcus$^{6}$, Ferdinand Kuemmeth$^{1}$\\}

\affiliation{ $^{1}$ Center for Quantum Devices, Niels Bohr Institute, University of Copenhagen, 2100 Copenhagen, Denmark\\
$^{2}$ Department of Physics and Astronomy, Birck Nanotechnology Center, Purdue University, West Lafayette, Indiana 47907, USA\\
$^{3}$ School of Materials Engineering and School of Electrical and Computer Engineering, Purdue University, West Lafayette, Indiana 47907, USA\\
$^{4}$ Department of Physics and Astronomy, Birck Nanotechnology Center, and Station Q Purdue, Purdue University, West Lafayette, Indiana 47907, USA\\
$^{5}$ School of Materials Engineering and School of Electrical and Computer Engineering, Purdue University, West Lafayette, Indiana 47907, USA\\
$^{6}$ Center for Quantum Devices and Station Q Copenhagen, Niels Bohr Institute, University of Copenhagen, 2100 Copenhagen, Denmark\\
$^{*}$ These authors contributed equally to this work}

\date{\today}

\begin{abstract}

By operating a one-electron quantum dot (fabricated between a multielectron dot and a one-electron reference dot) as a spectroscopic probe, we study the spin properties of a gate-controlled multielectron GaAs quantum dot at the transition between odd and even occupation number.  We observe that the multielectron groundstate transitions from spin-1/2-like to singlet-like to triplet-like as we increase the detuning towards the next higher charge state. The sign reversal in the inferred exchange energy persists at zero magnetic field, and the exchange strength is tunable by gate voltages and in-plane magnetic fields. Complementing spin leakage spectroscopy data, the inspection of coherent multielectron spin exchange oscillations provides further evidence for the sign reversal and, inferentially, for the importance of non-trivial multielectron spin exchange correlations. 

\end{abstract}

\maketitle

Semiconducting quantum dots with individual unpaired electronic spins offer a compact platform for quantum computation~\cite{Kloeffel_2013,Awschalom_2013}.
They provide submicron-scale two-level systems that can be operated as qubits~\cite{Loss_1998,Petta_2010,Bluhm_2011,Nowack_2007,Maune_2012,Malinowski_2016} 
and coupled to each other via direct exchange or direct capacitive interaction. In these approaches, the essential role of nearest-neighbor interactions in larger and larger arrays of one-electron quantum dots~\cite{Shulman_2012,Veldhorst_2015,Nichol_2016,Ito_2016,Zajac_2016} poses technological challenges to upscaling, due to the density of electrodes that define and control these quantum circuits. 
This issue has stimulated efforts to study long-range coupling of spin qubits either by electrical dipole-dipole interaction~\cite{Shulman_2012,Nichol_2016,Trifunovic_2012} or via superconducting microwave cavities~\cite{Mi_2017,Srinivasa_2016, RussPRB_2015}.
However, these approaches involve the charge degree of freedom, which makes the qubit susceptible to electrical noise~\cite{Coish_2005,Taylor_2007,Dial_2013,Martins_2016}. Recent work~\cite{Medford_2013_1, MalinowskiSRX_2017} indicates that the effective noise needs to be reduced significantly before long-range two-qubit gates with high fidelity can be reached~\cite{Srinivasa_2016,Russ_Topical_Review}.
Alternatively, symmetric exchange pulses can be implemented that perform fast, charge-insensitive gates~\cite{Weiss_2012,BertrandPRL2015,Martins_2016,Reed_2016}. 
Even though the exchange interaction is intrinsically short-ranged, its range can be increased by means of a quantum mediator~\cite{Baart_2016,Braakman_2013}. 
In particular, using a large multielectron quantum dot as an exchange mediator has the potential to do both: provide fast spin interaction~\cite{Mehl_2014,Srinivasa_2015} and alleviate spatial control line crowding. To avoid entanglement with internal degrees of freedom of the mediator, recent theory~\cite{Mehl_2014,Srinivasa_2015} motivates the use of a multielectron quantum dot with a spinless ground state and a level spacing sufficiently large to suppress unwanted excitations by gate voltage pulses.

\begin{figure}
	\begin{center}
	\includegraphics[width=0.48\textwidth]{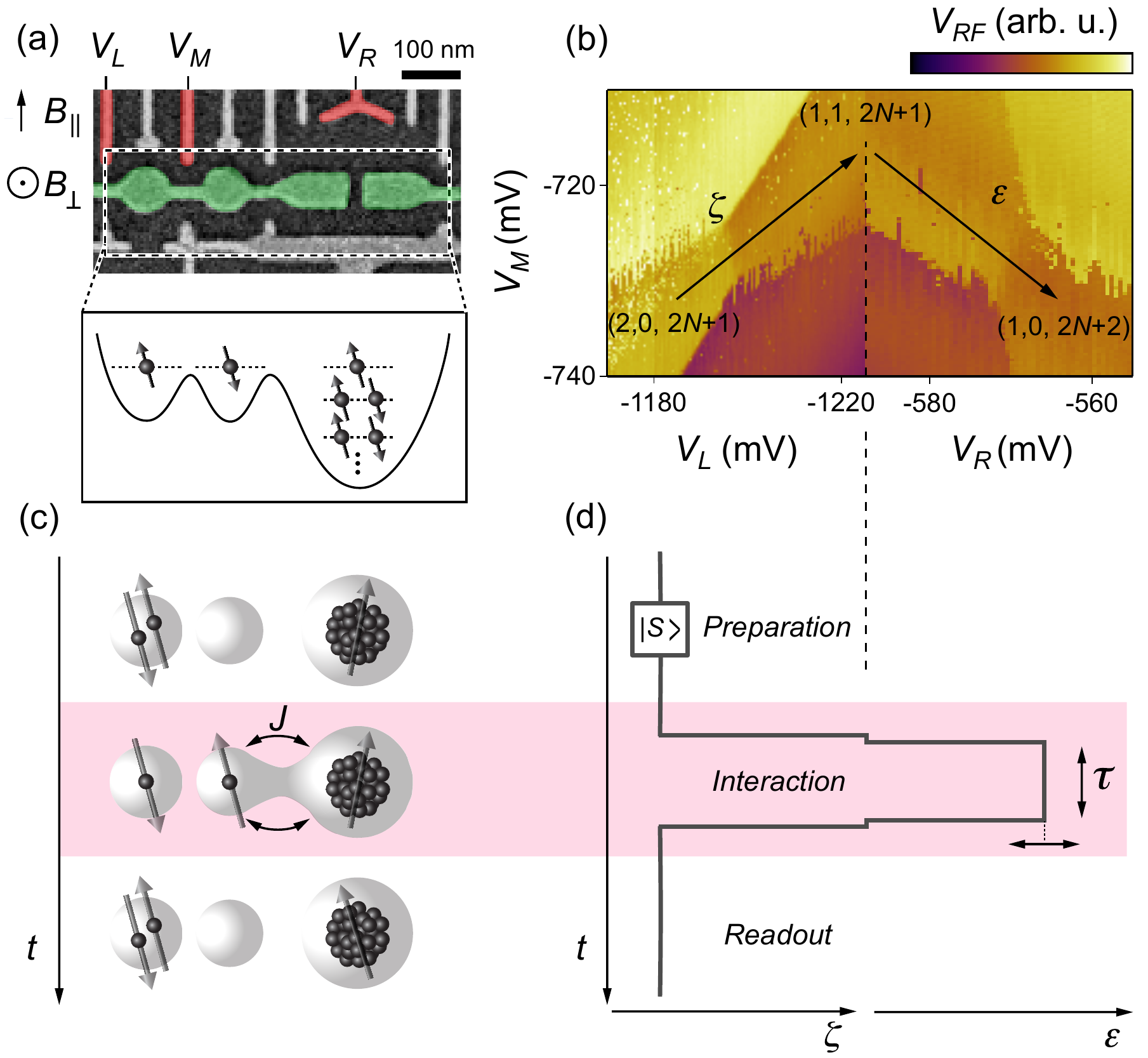}
	\caption{
	(a) Electron micrograph of the device consisting of a two-electron double quantum dot next to a multielectron quantum dot. The accumulation gate (colored in green) is operated at positive voltage. Remaining gates deplete the underlying two-dimensional electron gas. Gates $V_{L}$, $V_{M}$, and  $V_{R}$, highlighted in red, are connected to high-bandwidth lines. A proximal charge sensor (not shown) coupled to a radio frequency circuit allows fast measurements. The direction of the magnetic field $B_\parallel$ and $B_\perp$ is indicated.
	(b) Charge diagrams indicating the electron occupation of the triple quantum dot as function of $V_{L}$, $V_{M}$, and  $V_{R}$. Arrows indicate $\zeta$ and $\epsilon$ axes in a gate voltage space.
	(c) Concept of the experiment. Two electrons are initialized in a singlet state in the left quantum dot. Thereafter one of the electrons is moved to the middle dot and interacts with the multielectron quantum dot through exchange interaction $J$. At the end, readout is attained by performing spin-to-charge conversion for two-electron spin states in the double quantum dot.
	(d) Implementation of the pulse sequence in respect of the gate-voltage parameters $\zeta$ and $\varepsilon$.
	}
	\label{fig:fig1}
	\end{center}
\end{figure}

In this Letter, we investigate a GaAs multielectron quantum dot and show that its spin properties make it suitable for use as a coherent spin mediator. The experiment involves a chain of three quantum dots that can be detuned relative to each other using top-gate voltage pulses. 
The central one-electron dot serves as a probe: its spin can be tunnel coupled either to the left one-electron dot (serving as a reference spin for initialization and readout), or to a large dot on the right, thereby probing its multielectron spin states. 
We focus on a particular odd occupancy of the multielectron dot, $2N$+1, characterized by an effective spin 1/2, and establish that the exchange coupling between the central probe spin and the multielectron spin depends strongly and non-monotonically on the detuning of relevant gate voltages.  
Remarkably, this exchange coupling becomes negative, i.e. triplet-preferring, as the central electron is detuned further into the right dot. We therefore infer a spin-1 ground state for $2N$+2 occupation, even in the absence of an applied magnetic field. 
Besides fundamental implications for the role of non-trivial interactions within a multielectron dot, presented elsewhere for a large range of dot occupations~\cite{Malinowski_2017}, our finding has practical applications. 
For example, the nonmonotonicity of the exchange profile results in a sweetspot, whereas its sign reversal removes a long-standing constraint for the construction of compact dynamically corrected exchange gates~\cite{Wang_2012,Wang_2014}.

The three quantum dots were fabricated in a GaAs/Al$_{0.3}$Ga$_{0.7}$As heterostructure hosting a two-dimensional electron gas with a bulk density $n$~=~$2.5\times10^{15}$~m$^{-2}$ and a mobility  $\mu$~=~230~m$^{2}$/Vs, located 57 nm below the wafer surface.
The confining potential and dot occupancy is voltage-tuned by Ti/Au metallic gates deposited on a 10~nm thin HfO$_2$ gate dielectric. 
Figure~\ref{fig:fig1}(a) shows the two accumulation gates (colored in green) surrounded by various depletion gates, and a schematic cut through the resulting triple-well potential. Gates labeled $V_{L}$, $V_{M}$, and  $V_{R}$ (colored in red) are connected to high-bandwidth coaxial lines and allow application of nanosecond-scale voltage pulses. 
An adjacent quantum dot (not shown) serves as a fast charge sensor, i.e. changes in its conductance change the amplitude ($V_{RF}$) of a reflected rf carrier~\cite{Barthel_2010}.
All measurements were conducted in a dilution refrigerator with mixing chamber temperature below 30~mK. 

The device can be viewed as a two-electron double quantum dot (DQD) tunnel-coupled to a multielectron dot (MED) with an estimated number of electrons between 50 and 100, based on $n$ and the area of the multielectron dot.
By measuring $V_{RF}$ as a function of voltages $V_{L}$, $V_{M}$ and  $V_{R}$ we can map out the dots' occupancies in the vicinity of the charge states (2,0,$2N$+1), (1,1,$2N$+1) and (1,0,$2N$+2). Here, the numbers correspond to electron occupation in the left dot, central dot and the MED, respectively. 
The resulting charge diagram in Fig.~\ref{fig:fig1}(b) allows the definition of two detuning axes in gate-voltage space, $\zeta$ and $\varepsilon$, such that a reduction of $\zeta$ pushes the central electron into the left dot, whereas an increase in $\varepsilon$ pushes it to the MED (cf. arrows).

The MED spin states are probed by the pulse sequence illustrated in Fig.~\ref{fig:fig1}(c,d).
First, two electrons in a singlet state are prepared in the left dot, by pulsing to the (2,0,$2N$+1) charge state. 
Then a $\zeta$ pulse to the (1,1,$2N$+1) state effectively turns off intra-DQD exchange interactions while maintaining the two-electron spin state. 
The next step probes the interaction between the central electron and the MED in the vicinity of the charge transition between (1,1,$2N$+1) and (1,0,$2N$+2). 
This is done by pulsing $\varepsilon$, i.e. by temporarily applying a negative voltage pulse to $V_M$ and a positive voltage pulse to $V_R$.
After an interaction time $\tau$ we return to (1,1,$2N$+1) and immediately reduce $\zeta$ for single-shot reflectometry readout~\cite{Barthel_2009}: If $V_{RF}$ indicates a (2,0,$2N$+1) charge state, we assign a singlet outcome, whereas (1,1,$2N$+1) indicates that a spin interaction with the MED has occured, and we count it as a non-singlet outcome. The fraction of singlet outcomes when repeating typically 1024 identical pulse sequences is denoted by $P_S$. 

\begin{figure*}
	\centering
	\includegraphics[width=\textwidth]{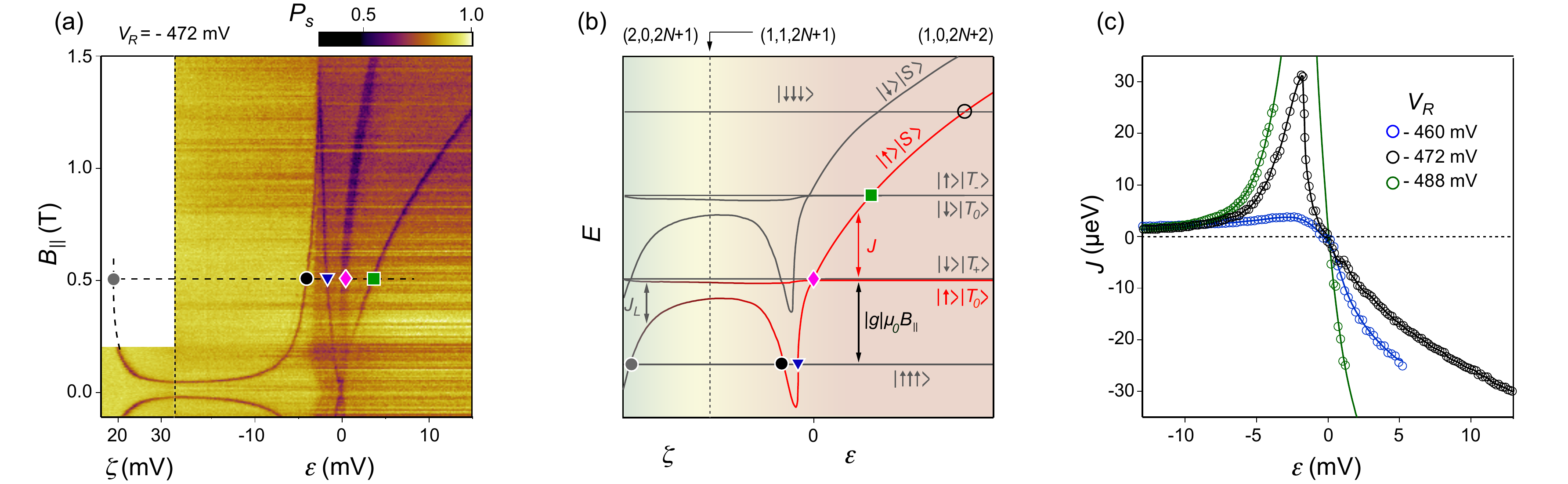}
	\caption{{
	(a) $P_S$ as a function of $\zeta$, $\varepsilon$ and $B_\parallel$ for a fixed, long interaction time $\tau=150$~ns. 
		(b) Corresponding energy diagram of the spin states of a Heisenberg model, as a function of $\zeta$, $\varepsilon$ for a fixed $B_\parallel$. 
	States highlighted in red witness the interaction between the central probe spin and the effective MED spin, which combine into a singlet-like state, $\ket{\uparrow}\ket{S}$, that is above a triplet-like state, $\ket{\uparrow}\ket{T_0}$, for sufficiently large $\varepsilon$ (negative $J$).  
	The charge character of the groundstate transitions from (2,0,2$N$+1) via (1,1,2$N$+1) to (1,0,2$N$+2) as indicated by the background shading. 
	The sign reversal of $J$ happens at $\varepsilon=0$.
	The Zeeman shift $|g| \mu_B B_\parallel$ and crossings with other states leading to spin leakage features in (a) are indicated (see main text). 
	Leakage from $\ket{\uparrow}\ket{S}$ to the fully polarized $\ket{\downarrow\downarrow\downarrow}$ state (empty circle) is not observed in (a), likely because weak Overhauser gradients or spin-orbit coupling do not allow changes in spin projection by 2. 
	 The $S$-$T_0$ leakage feature in (a) (magenta diamond) is not field independent as predicted by the model, likely due to orbital coupling of $B_\parallel$ to MED states in combination with a small misalignment of the sample.
	(c) Experimental exchange profiles for different operating points (distortions of the confining potential), identified by $V_{R}$ during the readout step (symbols). Black circles are extracted from (a). Solid lines are guides to the eye.
}}
\label{fig:fig2}
\end{figure*}

Leakage spectroscopy is performed by choosing  $\tau$ sufficiently long to detect incoherent spin mixing between the central electron and MED states.  
Figure~\ref{fig:fig2}(a) shows $P_S(\varepsilon, B_\parallel)$, where $\varepsilon$ is the detuning voltage during the interaction step and  $B_\parallel$ is the applied in-plane magnetic field. To make connection to the conventional two-electron DQD regime we also plot $P_S(\zeta, B_\parallel)$, acquired by replacing the composite $\zeta$-$\varepsilon$ pulse in Fig.~\ref{fig:fig1}(d) by a pure  $\zeta$ pulse. 
Spin leakage is clearly observed as a sharp suppression of $P_S$ for particular detuning values, with a non-trivial magnetic field  dependence for $\varepsilon>-5$~mV. To understand this spectrum we note that all features below $\varepsilon\approx-5$~mV are well explained by mixing with fully polarized spin states, consistent with previous spin leakage experiments: The $\varepsilon$ dependence (``spin funnel'') is analogous to mixing between singlet and $T_+\equiv\ket{\uparrow\uparrow}$ in two-electron DQDs~\cite{Petta_2005,Maune_2012,Veldhorst_2015}, whereas the $\zeta$ dependence is analogues to mixing between a singlet-like state and $\ket{\uparrow\uparrow\uparrow}$ in three-electron triple quantum dots~\cite{Medford_2013_1,Medford_2013_2}. 
(Here, each arrow indicates the spin state within one quantum dot.)
The characteristic dependence on $B_\parallel$ arises from the linear Zeeman shift of fully polarized spin states~\cite{Laird_2010,Taylor_2013}.

This identification confirms odd multielectron occupation, i.e. (1,1,$2N$+1), with effective spin 1/2. It also allows us to extract the exchange interaction ($J$) between the central spin and the effective MED spin from the ordinate of the leakage feature (black dot), using  $J = |g| \mu_B B_\parallel$, where $g = -0.44$ is the electronic $g$-factor for GaAs and $\mu_B$ is the Bohr magneton. 
Towards higher detuning, $\varepsilon>-5$~mV, an overall drop in the background of $P_S$ indicates that the MED ground state transitions into (1,0,$2N$+2), approximately concurrent with the sharp leakage feature (black dot) reaching a maximum before turning towards $B_\parallel=0$ (blue triangle). At $\varepsilon=0$ two additonal leakage features appear at $B_\parallel=0$. We interpret this maximum as a maximum in in the exchange profile, $J(\varepsilon)$, and associate the crossing at $B_\parallel=0$ with a sign reversal of $J(\varepsilon)$.

To infer the spin spectrum, we impose the observed exchange profile $J(\zeta,\varepsilon)$ on a Heisenberg model of three spin-1/2 orbitals~\cite{footnoteFK}. For simplicity we ignore orbital coupling to $B_\parallel$ and inspect spin Zeeman effects only.
The resulting energy diagram, sketched in Fig.~\ref{fig:fig2}(b) for finite $B_\parallel$, allows us to identify all characteristic leakage features.
On the left side of the energy diagram only intra-DQD exchange is significant ($J_{L}$), and the eigenstates are the tensor products of a DQD spin state and a MED ``spectator" spin. For example, the grey dot marks the crossing between $\ket{S}\ket{\uparrow}$ and $\ket{T_+}\ket{\uparrow}$, and relates the ``spin funnel'' in (a) to $J_{L}(\zeta)$.
Analogously, on the right side of the energy diagram, the left dot is decoupled and hosts the spectator spin, while the central spin interacts with the effective MED spin. Here, field dependent crossings map out the positive (black and blue marker) and negative (green marker) regime of $J(\varepsilon)$ (cf. crossings of $\ket{\uparrow}\ket{S}$ with $\ket{\uparrow\uparrow\uparrow}$, $\ket{\uparrow}\ket{T_0}$ and $\ket{\downarrow}\ket{T_+}$). 
At these crossings rapid mixing due to uncontrolled Overhauser gradients is expected to occur, changing electronic spin projections by 1 on a timescale of $T_2^*\approx10$ ns \cite{Petta_2010}.

\begin{figure}[b]
	\begin{center}
	\includegraphics[width=0.48\textwidth]{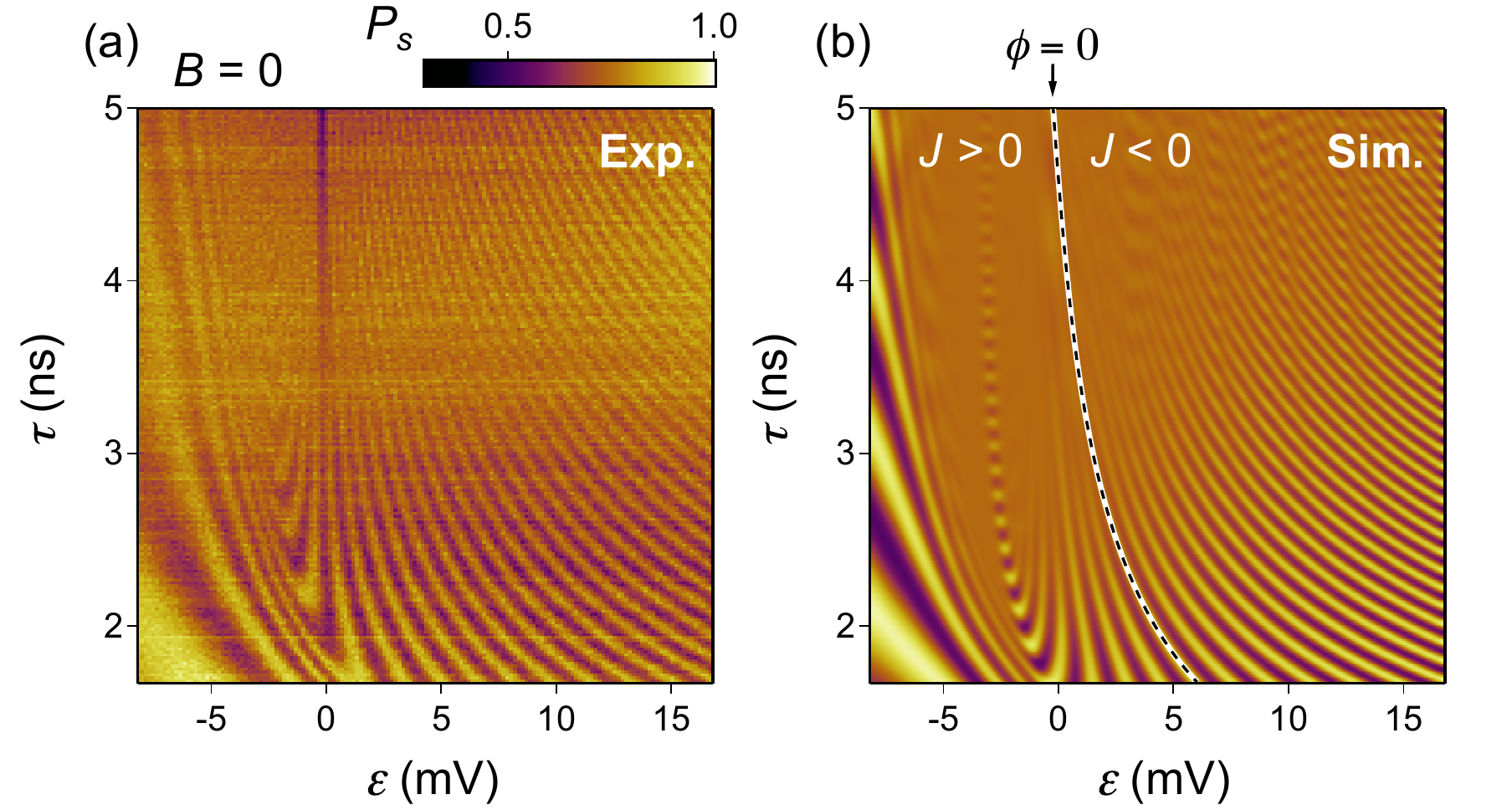}
	\caption{
	(a) Exchange oscillations in $P_S$ as a function of $\varepsilon$ and exchange time $\tau$, in the vicinity of (1,1,$2N$+1) and (1,0,$2N$+2) charge transition. External magnetic field is zero, and DC tuning voltages are the same as in Fig.~\ref{fig:fig2}(a) ($V_{R}$~=~472~mV).
	(b) Simulation of the exchange oscillations using $J(\varepsilon)$ from Fig.~\ref{fig:fig2}(c). Simulation assumes a Gaussian $\varepsilon$ low frequency noise with a standard deviation of 0.18 mV, and a rise time of the experimental instrumentation of 0.8~ns. Dashed line indicates zero phase accumulation and divides the area where $J$ is positive and negative. For large $\tau$ a dark feature appears at $\varepsilon = 0$ in (a) but not in (b). We associate it with leakage out of the simulated subspace, corresponding to the crossing in Fig.~\ref{fig:fig2}(a,b) indicated by a green square. 
}
	\label{fig:fig3}
	\end{center}
\end{figure}

In contrast to three-electron triple dots~\cite{Medford_2013_1,Medford_2013_2}, where $J$ is always positive, we observe that $\ket{\uparrow}\ket{S}$ and $\ket{\uparrow}\ket{T_0}$ cross each other at $\varepsilon=0$.
This implies that the exchange interaction between the single and multielectron quantum dot changes sign from positive to negative, i.e. it is singlet-preferring for small hybridization and becomes triplet-preferring once the central electron has transferred to the multielectron dot. Next, we test for robustness and gate-tunability of this effect.
In Fig.~\ref{fig:fig2}(c) we plot $J(\varepsilon)$ extracted from Fig.~\ref{fig:fig2}(a) (black symbols), and compare it to two exchange profiles (green and blue symbols) measured by distorting the confining potential while preserving the charge configuration of the triple dot system (cf. Supplementary Fig.~S1).
In all cases $J(\varepsilon)$ shows the same behavior, namely a maximum and sign reversal at the position of the charge transition, and a negative sign in the (1,0,$2N$+2) configuration.
This interpretation implies that the $2N$+2 charge state of the multielectron dot has a total spin of 1 at zero magnetic field, which is further confirmed by studying the MED behavior over multiple charge states~\cite{Malinowski_2017}.

Direct evidence for the sign reversal in $J$ (without the need for a magnetic field) can be obtained from time-domain measurements. To this end, we induce coherent exchange oscillations between central and MED spin by significantly reducing (and varying) the interaction time $\tau$. The observed pattern of $P_S(\varepsilon,\tau)$, shown in Fig.~\ref{fig:fig3}(a) for the same DC tuning parameters as in Fig.~\ref{fig:fig2}(a), differ from analogous oscillations of the exchange-only qubit \cite{Laird_2010,Medford_2013_2}. 
Namely, the appearance of a chevron-like pattern indicates the presence of a local maximum in $J(\varepsilon)$. 
Following contours of equal phase ($\phi$) around this ``sweet spot'', we note that $\phi(\tau)$ has opposite sign for large and small $\varepsilon$, implying a sign reversal in $J(\varepsilon)$. 
To show consistency between time-domain and leakage spectroscopy results, we perform numerical simulations of the exchange oscillations using the experimentally measured exchange profile presented in Fig.~\ref{fig:fig2}(c). The simulation is limited to the Hilbert space spanned by $\ket{\uparrow}\ket{S}$ and $\ket{\uparrow}\ket{T_0}$ (indicated with red in Fig.~\ref{fig:fig2}(b)) and includes a quasistatic Gaussian noise in $\varepsilon$ with standard deviation $\sigma_{\epsilon}$~=~0.18~mV~\cite{Martins_2016,Barnes_2016} and a rise time of our instrumentation of 0.8~ns. The simulation reproduces a chevron pattern (Fig.~\ref{fig:fig3}(b)), whereas simulations using $J(\varepsilon)=|J(\varepsilon)|$ produce a qualitatively different pattern (not shown). Therefore, the contour $\phi=0$ does indeed separate regions with $J>0$ from regions with $J<0$.

\begin{figure}
	\begin{center}
	\includegraphics[width=0.48\textwidth]{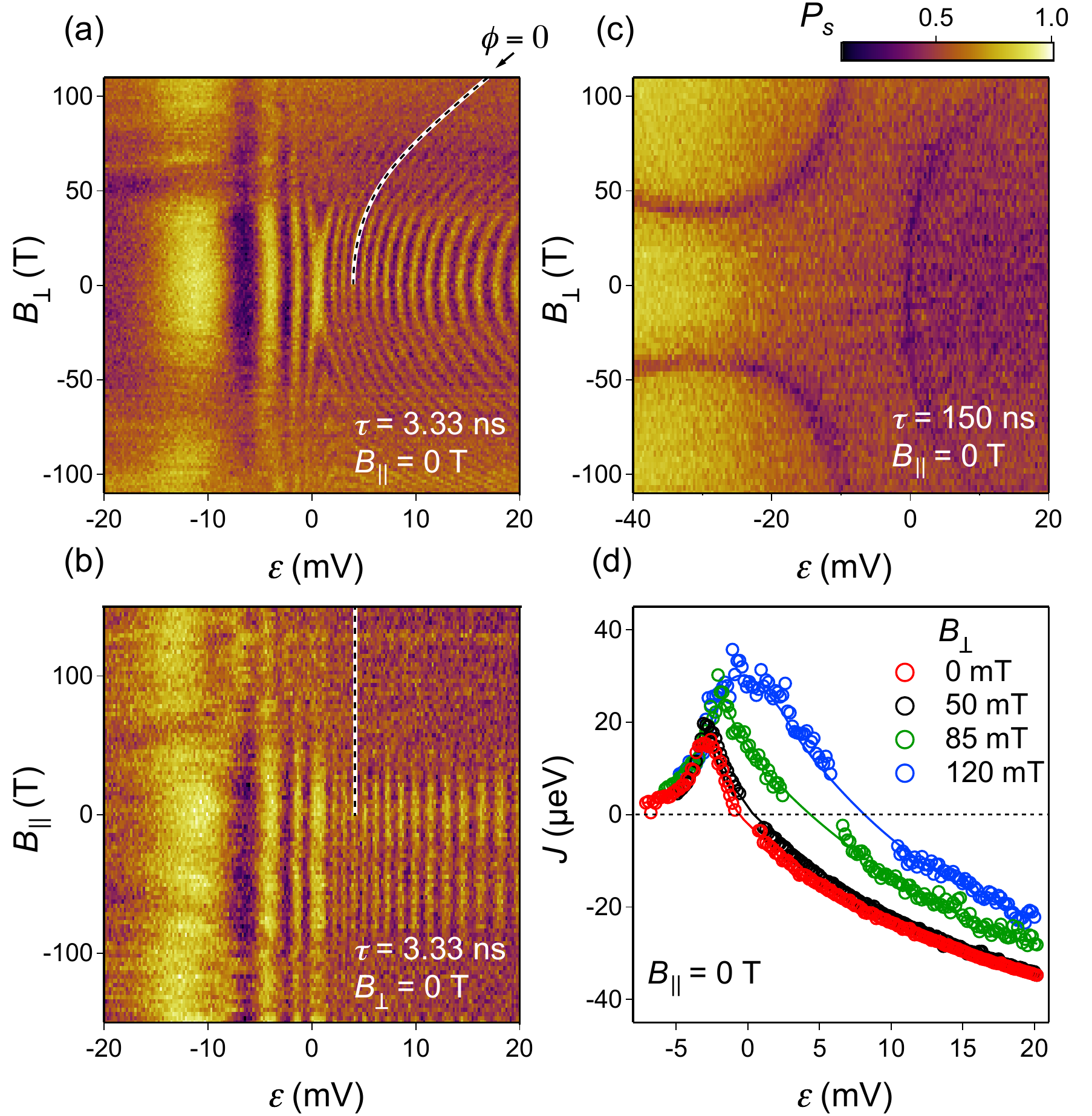}
	\caption{
	(a) Exchange oscillations as a function of orbital field $B_\perp$ and $\varepsilon$ for fixed exchange time $\tau$~=~3.33~ns and fixed $B_\parallel$~=~0~T. 
	(b) Exchange oscillations as a function of $B_\parallel$ and $\varepsilon$ for fixed exchange time $\tau$~=~3.33~ns and fixed $B_\perp$~=~0~T.
	(c) Same as (a) but in the leakage spectroscopy regime ($\tau$~=~150~ns). Features of reduced $P_S$ correspond to mixing between $\ket{\uparrow}\ket{S}$ and various other states (cf. horizontal cut of Fig.~\ref{fig:fig2}(a) at $B_\parallel=0$). A small deviation between the $J=0$ feature in (c) and the $\phi=0$ contour in (a) is likely due the drastically different choices of $\tau$ in combination with finite-rise-time effects of our instrumentation. 
	(d) Exchange profiles $J(\varepsilon)$ for different values of $B_\perp$.
	}
	\label{fig:fig4}
	\end{center}
\end{figure}

Finally, we study the effects of applied magnetic fields on the exchange profile. 
Figure~\ref{fig:fig4}(a) presents $P_S$ as a function of $\varepsilon$ and out-of-plane magnetic field $B_\perp$, while keeping $B_\parallel$~=~0 and $\tau$~=~3.33~ns fixed. 
In such a plot, contours correspond to constant $J$ in the $\varepsilon$-$B_\perp$ plane, and their curvature indicates that out-of-plane magnetic fields move the sign reversal of $J$ towards higher detuning (cf. $\phi=0$ contour, marked by a dashed line). 
For comparison, within the same range, $B_\parallel$ has no observable influence on the pattern of the exchange oscillations (Fig.~\ref{fig:fig4}(b)). 
By choosing $\tau$ longer than the coherence time we obtain the $B_\perp$-dependence of the leakage spectrum (Fig.~\ref{fig:fig4}(c), using $\tau$~=~150~ns). 
The two leakage features appearing for negative values of $\varepsilon$ correspond to mixing between $\ket{\uparrow}\ket{S}$ and the fully polarized $\ket{\uparrow\uparrow\uparrow}$. The leakage feature appearing for positive values of $\varepsilon$ indicates $J=0$ and resolves into three lines at higher fields (cf. Fig.~\ref{fig:fig2}). 

Exchange profiles $J(\varepsilon)$ for $B_\perp$ =  0, 50, 85 and 120 mT were extracted from $P_S(\varepsilon,\tau)$ maps obtained for the same tuning voltages as in Fig.~\ref{fig:fig4}(a) (Supplementary Fig.~S2). Their $B_\perp$-dependence shown in Fig.~\ref{fig:fig4}(d) corroborates again the sensitivity of the exchange profile to the underlying electronic orbitals, and establishes an electrical  sweet spot in $J(\varepsilon)$ that can be precisely tuned by $B_\perp$. 

In summary, we have investigated experimentally the exchange interaction between a two-electron double quantum dot and a multielectron quantum dot, by complementing incoherent spin leakage measurements with time-resolved coherent exchange oscillations at various tuning voltages and magnetic field configurations. 
We find that the multielectron dot with odd occupation number behaves as a spin-1/2 object that gives rise to a non-monotonic exchange coupling to the neighboring dot. By changing the relative dot detuning voltage by a few millivolt the sign of the exchange interaction can be tuned from positive to negative (also at zero magnetic field), indicating the presence of non-trivial electron-electron interactions. Finally, we show that the exchange profile can be tuned by either changing the gate potentials or applying an out-of-plane orbital magnetic field, giving rise to a tunable electrical sweet spot that might benefit the implementation of high-fidelity exchange gates~\cite{Martins_2016,Reed_2016} in long-distance quantum mediators.

We thank Stephen Bartlett, Andrew Doherty and Thomas Smith for helpful discussions.
This work was supported by LPS-MPO-CMTC, the EC FP7-ICT project SiSPIN no. 323841, the Army Research Office and the Danish National Research Foundation.

\clearpage
\onecolumngrid
\setcounter{figure}{0}
\setcounter{equation}{0}
\setcounter{page}{1}

\renewcommand{\thefigure}{S\arabic{figure}}  
\renewcommand{\theequation}{S\arabic{equation}}
\renewcommand{\thetable}{S\arabic{table}}
\renewcommand{\bibnumfmt}[1]{[S#1]}
\renewcommand{\citenumfont}[1]{S#1}

\section{Supplemental Material:}
\section{Negative spin exchange in a multielectron quantum dot}

\begin{figure}[!hbt]
\begin{center}
\includegraphics[width=110 mm]{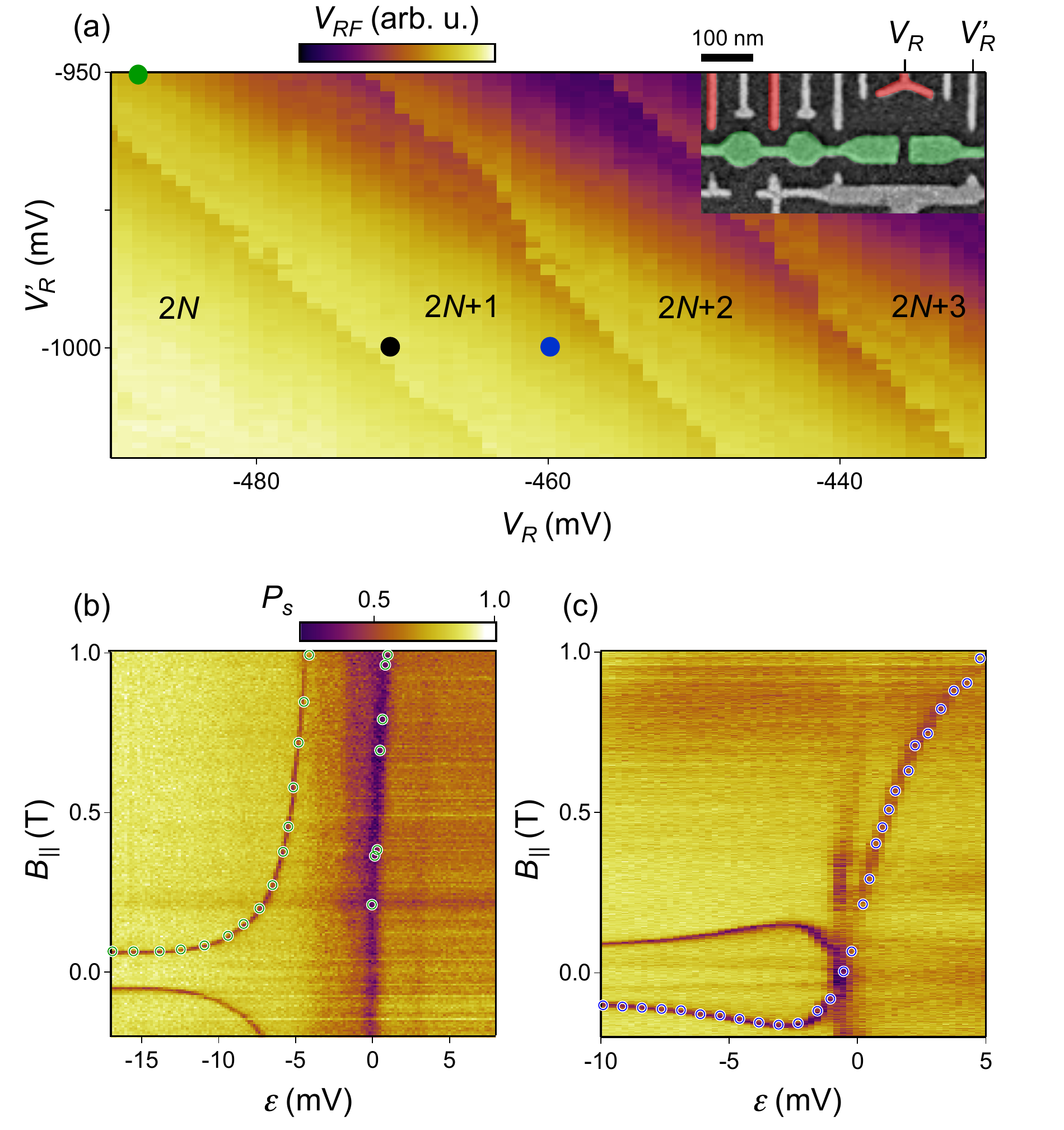}
\caption{
(a) Charge diagram indicating the electron occupation in the multielectron dot as function of gates $V_{R}$ and  $V_{R}'$, as defined in the micrograph shown in the inset. Dots indicate the DC values of $V_{R}$ and $V_{R}'$ at which spectroscopy of the exchange energy have been performed.
(b) Probability of detecting a singlet, $P_{s}$, as a function of $\varepsilon$ and $B_{||}$ for  a exchange time $\tau=150$~ns, $V_{R}$~=~-490~mV and $V_{R}'$~=~-950~mV.
(c) Same as (b) for $V_{R}$~=~-460~mV and $V_{R}'$~=~-1000~mV.
Data corresponding to the black dot is shown in the main article in Fig. 2(a).
Exchange profile, $J$, extracted from these two spectroscopies is shown in Fig. 2(c).
}
\label{fig:Sfig1}
\end{center}
\end{figure}

\begin{figure}
\begin{center}
\includegraphics[width=150 mm]{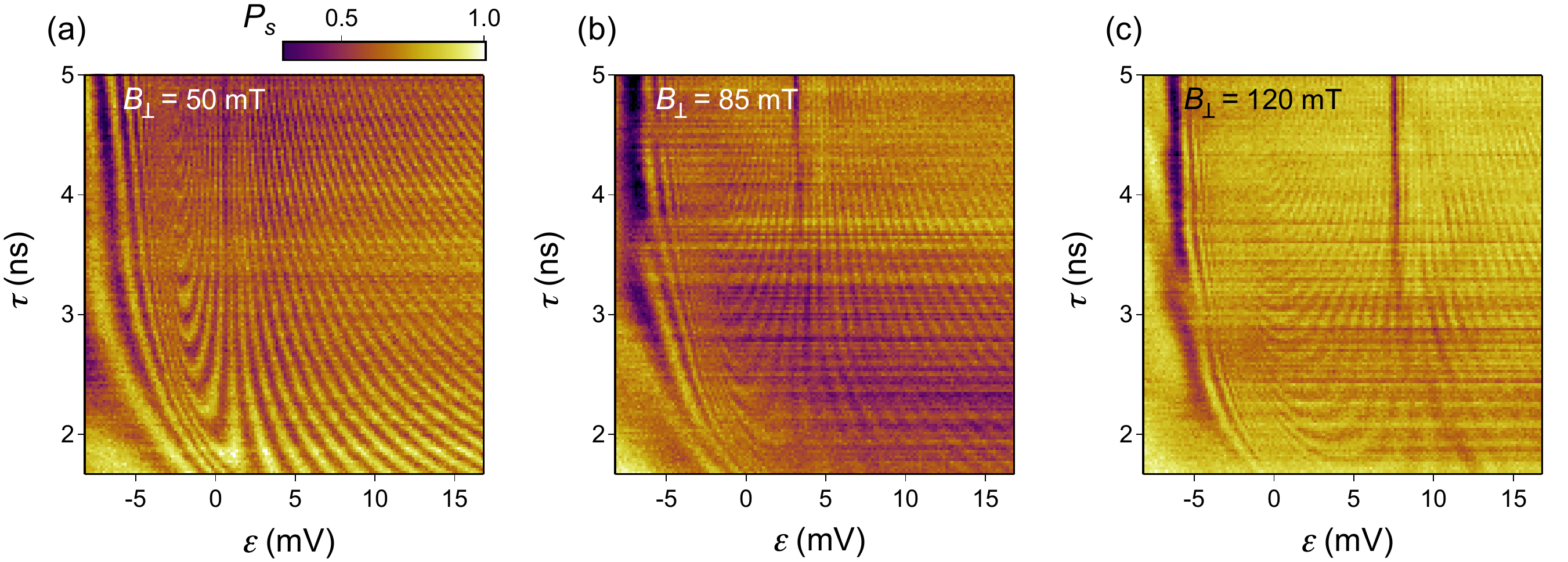}
\caption{ 
Exchange oscillations as a function of $\varepsilon$ for a exchange time $\tau$, in the vicinity of the (1,1,$N$)-(1,0,$N$+1) charge transition for various values of $B_\perp$. 
$J$ profiles extracted from these oscillations are plotted in  Fig. 4(d) in the main text.
(a) $B_\perp =50$~mT; (b) $B_\perp =85$~mT;  (c) $B_\perp =120$~mT.
}
\label{fig:Sfig2}
\end{center}
\end{figure}

\subsection{Extracting $J(\varepsilon)$ from exchange oscillations}

Exchange profiles $J(\varepsilon)$ plotted in Fig.~4(d) in the main text were obtained from Fig.~3(a) and Figs.~\ref{fig:Sfig2}(a-c) for $B_\perp $~=~0, 50, 85 and 120~mT, respectively.
For each value of $\varepsilon$, the frequency of the exchange oscillations $J$ is obtained in two steps. 
First, we calculate the Fast Fourier transform of $P_\mathrm{s}(\tau)$ and find the frequency bin with the largest weight. 
Then we use this frequency as an initial guess for fitting a damped sine wave of frequency $J$ to $P_\mathrm{s}(\tau)$, with a decaying amplitude of the form exp$\left(-\tau/T_\mathrm{R}\right)$. 
Values of $J(\varepsilon)$ extrated by this method are plotted as circles in Fig.~4(d).

\end{document}